\documentclass[12pt,thmsa]{article}
\usepackage{amsfonts}

\usepackage{sw20aip}

\input{tcilatex}
\begin{document}

\author{Ron Rubin and Nathan Salwen \\
Lyman Laboratory of Physics, Harvard University\\
Email: rubin@string.harvard.edu}
\title{A Parity-Conserving Canonical Quantization for the Baker's Map}
\date{January 22, 1998}
\maketitle

\begin{abstract}
We present here a complete description of the quantization of the baker's
map. The method we use is quite different from that used in Balazs and Voros
[BV] and Saraceno [S]. We use as the quantum algebra of observables the
operators generated by $\left\{ \exp \left( 2\pi i\widehat{x}\right) ,\exp
\left( 2\pi i\widehat{p}\right) \right\} $ and construct a unitary
propagator such that as $\hbar \rightarrow 0$, the classical dynamics is
returned. For Planck's constant $h=1/N$, we show that the dynamics can be
reduced to the dynamics on an $N$-dimensional Hilbert space, and the unitary 
$N\times N$ matrix propagator is the same as given in [BV] except for a
small correction of order $h$. This correction is is shown to preserve the
symmetry $x\rightarrow 1-x$ and $p\rightarrow 1-p$ of the classical map for
periodic boundary conditions.
\end{abstract}

\subsection{The Classical Baker's Map and its Covering Map}

\smallskip The classical baker's map is a mapping of the unit torus onto
itself defined as follows. Let $x$ and $p$ be the coordinates on the torus
and take 
\begin{equation}
(x,p)\rightarrow (x^{\prime },p^{\prime })=\left\{ 
\begin{array}{ll}
(2x,p/2), & 0\leq x<1/2; \\ 
(2x-1,p/2+1/2), & 1/2\leq x<1.
\end{array}
\right.  \label{Classical baker's map}
\end{equation}

This map describes a stretching in $x$, shrinking in $p$, and chopping and
stacking, similar to the way bakers make certain pastries. The motion on the
torus is completely chaotic with a positive Liapunov exponent $\log 2$, and
in fact is a paradigm for the study of classical chaos. For more details on
the classical baker's map, including a description of the map as a dynamics
on binary digits, we refer the reader to [BV].

A quantum version of the map was introduced by Balazs and Voros [BV] and
then by Saraceno [S]. In this description, the dynamics is quantized for
values of Planck's constant satisfying $h=1/N$, by constructing a quantum
propagator $u_{mn}$ as a unitary $N\times N$ matrix. The classical limit is
demonstrated numerically for $N\rightarrow \infty $. Below we present a
quantum propagator which ``reduces'' to the finite dimensional matrix
propagator given in [BV] at the point $\theta =\left( 0,0\right) $,
corresponding to periodic boundary conditions, except for a correction of
order $\hbar $. This correction is what preserves the classical symmetry
broken in the quantization scheme presented in [BV]. At other points on the $%
\theta $-torus, the question of whether the propagator has a
finite-dimensional ``fixed point'' remains open. It is the author's guess
that this is the case, and it remains an interesting and fairly
straightforward extension of this work.

Our quantization procedure uses the fact that the baker's map has a natural
lift to the universal covering $\mathbb{R}^{2}$of $\mathbb{T}^{2}$ given by 
\begin{equation}
\beta :(x,p)\rightarrow (x^{\prime },p^{\prime })=\left\{ 
\begin{array}{ll}
(2x,p/2), & (x,p)\in l\cap e_{p}; \\ 
(2x-1,p/2+1/2), & (x,p)\in r\cap e_{p}; \\ 
(2x+1,p/2+1/2), & (x,p)\in l\cap o_{p}; \\ 
(2x,p/2), & (x,p)\in r\cap o_{p},
\end{array}
\right.  \label{coveringmap}
\end{equation}
so that for $a,b\in \mathbb{Z}$, 
\begin{equation}
\beta ^{*}e^{2\pi i\left( ax+bp\right) }=e^{4\pi iax}e^{i\pi p}\left( \chi
_{l}\left( x\right) +\left( -1\right) ^{b}\chi _{r}\left( x\right) \right)
\left( \chi _{e_{p}}\left( p\right) +\left( -1\right) ^{b}\chi
_{o_{p}}\left( p\right) \right) ,  \label{hope}
\end{equation}
where 
\begin{eqnarray}
l &:&=\left\langle [0,1/2)+\mathbb{Z}\right\rangle \times \mathbb{R}, 
\nonumber \\
r &:&=\left\langle [1/2,1)+\mathbb{Z}\right\rangle \times \mathbb{R},
\label{regone} \\
e_{p} &:&=\mathbb{R}\mathbf{\times }\left\langle [0,1)+2\mathbb{Z}%
\right\rangle ,  \nonumber \\
o_{p} &:&=\mathbb{R}\mathbf{\times }\left\langle [1,2)+2\mathbb{Z}%
\right\rangle .  \nonumber
\end{eqnarray}
and 
\begin{eqnarray*}
\chi _{l}(x) &=&\left\{ 
\begin{array}{ll}
1, & x\in [0,1/2)+\mathbf{Z} \\ 
0, & otherwise
\end{array}
\right. , \\
\chi _{r}(x) &=&\left\{ 
\begin{array}{ll}
1, & r\in [1/2,1)+\mathbf{Z} \\ 
0, & otherwise
\end{array}
\right. , \\
\chi _{e_{p}}(p) &=&\left\{ 
\begin{array}{ll}
1, & p\in [0,1)+2\mathbf{Z} \\ 
0, & otherwise
\end{array}
\right. , \\
\chi _{o_{p}}(p) &=&\left\{ 
\begin{array}{ll}
1, & p\in [1,2)+2\mathbf{Z} \\ 
0, & otherwise
\end{array}
\right. .
\end{eqnarray*}
The inverse of this map is as follows: 
\begin{equation}
\beta ^{-1}:(x,p)\rightarrow (x^{\prime },p^{\prime })=\left\{ 
\begin{array}{ll}
(x/2,2p), & (x,p)\in e_{x}\cap b; \\ 
(x/2-1/2,2p-1), & (x,p)\in o_{x}\cap b; \\ 
(x/2+1/2,2p-1), & (x,p)\in e_{x}\cap t; \\ 
(x/2,2p), & (x,p)\in o_{x}\cap t,
\end{array}
\right.  \label{inversecoveringmap}
\end{equation}
where we have used the ``conjugate'' regions 
\begin{eqnarray}
b &:&=\mathbb{R}\mathbf{\times }\left\langle [0,1/2)+\mathbb{Z}\right\rangle
,  \nonumber \\
t &:&=\mathbb{R}\mathbf{\times }\left\langle [1/2,1)+\mathbb{Z}\right\rangle
,  \label{regtwo} \\
e_{x} &:&=\left\langle [0,1)+2\mathbb{Z}\right\rangle \mathbf{\times }%
\mathbb{R},  \nonumber \\
o_{x} &:&=\left\langle [1,2)+2\mathbb{Z}\right\rangle \times \mathbb{R}. 
\nonumber
\end{eqnarray}
Observe that these subsets of $\mathbb{R}^{2}$ satisfy the following
relations: 
\begin{eqnarray*}
l\cup r &=&b\cup t=e_{p}\cup o_{p}=e_{x}\cup o_{x}=\mathbb{R}^{2}, \\
l\cap r &=&b\cap t=e_{p}\cap o_{p}=e_{x}\cap o_{x}=\mathbf{\emptyset .}
\end{eqnarray*}

\subsection{The Quantum Propagator}

The outline of our quantization can now be described as follows. The quantum
algebra of observables is restricted to the set of operators generated by $%
\left\{ U=\exp \left( 2\pi i\widehat{x}\right) ,V=\exp \left( 2\pi i\widehat{%
p}\right) \right\} \,$(the quantization of the classical algebra of periodic
functions). We construct a quantum propagator by quantizing the dynamics of
the covering map (\ref{coveringmap}). The quantum dynamics induced on the
algebra of observables for the quantum torus is the quantum baker's map. In
this sense, this quantization is similar to one given in [DEG].

We now construct the quantum propagator $F$. The kinematics is already
given: the Hilbert space is the usual $L^{2}(\mathbb{R})$. For the dynamics,
we work in the Heisenberg picture, and first give the quantum analogs of
equations (\ref{regone}) and (\ref{regtwo}). We define the following
projection operators: 
\begin{eqnarray}
L &:&=\int_{[0,1/2)+\mathbb{Z}}\left| x\right\rangle \left\langle x\right|
\;dx,  \label{Proj1} \\
R &:&=\int_{[1/2,1)+\mathbb{Z}}\left| x\right\rangle \left\langle x\right|
\;dx,  \nonumber \\
B &:&=\int_{[0,1/2)+\mathbb{Z}}\left| p\right\rangle \left\langle p\right|
\;dp,  \nonumber \\
T &:&=\int_{[1/2,1)+\mathbb{Z}}\left| p\right\rangle \left\langle p\right|
\;dp,  \nonumber
\end{eqnarray}
and 
\begin{eqnarray}
E_{x} &:&=\int_{[0,1)+2\mathbb{Z}}\left| x\right\rangle \left\langle
x\right| \;dx,  \label{Proj2} \\
O_{x} &:&=\int_{[1,2)+2\mathbb{Z}}\left| x\right\rangle \left\langle
x\right| \;dx,  \nonumber \\
E_{p} &:&=\int_{[0,1)+2\mathbb{Z}}\left| p\right\rangle \left\langle
p\right| \;dp,  \nonumber \\
O_{p} &:&=\int_{[1,2)+2\mathbb{Z}}\left| p\right\rangle \left\langle
p\right| \;dp.  \nonumber
\end{eqnarray}
Observe that 
\[
L+R=B+T=E_{x}+O_{x}=E_{p}+O_{p}=I, 
\]
and 
\[
LR=BT=E_{x}O_{x}=E_{p}O_{p}=0. 
\]
We next define appropriate ``shift'' operators. A shift in $p,$ or a shift
in $x$, by unity is achieved by the following unitary operators,
respectively: 
\begin{eqnarray*}
X &=&e^{i\widehat{x}/\hbar }, \\
Y &=&e^{i\widehat{p}/\hbar }.
\end{eqnarray*}
Note that $X$ and $Y$ commute with the algebra $\frak{A}_{\hbar }$ generated
by $U$ and $V$. We shall also need the following commutation relations: 
\begin{eqnarray}
XL &=&LX,\quad YL=LY,\quad XR=RX,\quad YR=RY,  \label{commutation relations}
\\
Y^{1/2}L &=&RY^{1/2},\quad X^{1/2}B=TX^{1/2},  \nonumber \\
YE_{x} &=&O_{x}Y,\quad XE_{p}=O_{p}X.  \nonumber
\end{eqnarray}
We demonstrate one of these commutation relations explicitly. The others
involve similar calculations: 
\begin{eqnarray*}
YL &=&e^{i\widehat{p}/\hbar }\int_{[0,1/2)+\mathbb{Z}}\left| x\right\rangle
\left\langle x\right| \;dxe^{-i\widehat{p}/\hbar }e^{i\widehat{p}/\hbar } \\
&=&\int_{[0,1/2)+\mathbb{Z}}\left| x-1\right\rangle \left\langle x-1\right|
\;dxe^{i\widehat{p}/\hbar } \\
&=&\int_{[0,1/2)+\mathbb{Z}}\left| x\right\rangle \left\langle x\right|
\;dxe^{i\widehat{p}/\hbar }=LY.
\end{eqnarray*}
We next find the unitary operator $S$ which takes $\widehat{x}$ to $2%
\widehat{x}$ and $\widehat{p}$ to $\widehat{p}/2$. We construct this
operator by intuition and appealing to the corresponsding classical action
which takes $x\rightarrow 2x$ and $p\rightarrow p/2$. Consider the
commutator of $\widehat{x}$ with $\left( \widehat{x}\widehat{p}+\widehat{p}%
\widehat{x}\right) ^{n}$. We find for any integer $n$%
\[
\widehat{x}\left( \widehat{x}\widehat{p}+\widehat{p}\widehat{x}\right)
^{n}=\left( \widehat{x}\widehat{p}+\widehat{p}\widehat{x}+2i\hbar \right)
^{n}\widehat{x} 
\]
thus formally, for any operator expandable as a Taylor series in $\widehat{x}%
\widehat{p}+\widehat{p}\widehat{x}$, we find 
\[
\widehat{x}f\left( \widehat{x}\widehat{p}+\widehat{p}\widehat{x}\right)
=f\left( \widehat{x}\widehat{p}+\widehat{p}\widehat{x}+2i\hbar \right) 
\widehat{x}. 
\]
Thus, if we define the operator 
\begin{equation}
S:={\large \exp }\left( -\frac{i\log 2}{2\hbar }(\widehat{x}\widehat{p}+%
\widehat{p}\widehat{x})\right)  \label{stretching}
\end{equation}
we see formally 
\begin{eqnarray*}
\widehat{x}S &=&2S\widehat{x}, \\
\widehat{p}S &=&S\widehat{p}/2,
\end{eqnarray*}
or 
\begin{eqnarray*}
S^{\dagger }\widehat{x}S &=&2\widehat{x}, \\
S^{\dagger }\widehat{p}S &=&\widehat{p}/2.
\end{eqnarray*}
We can make this argument rigorous with straightforward continuity argument,
which we omit here. Observe also that the operator $S$ is unitary since $%
\widehat{x}\widehat{p}+\widehat{p}\widehat{x}$ is Hermitian.

We are now in a position to write down a propagator for the baker's map.
Based on equation \ref{hope}, we have the following definition.

\begin{definition}
\textbf{(Baker's Map Propagator)} Let the operator $F$ be defined as
follows: 
\begin{eqnarray}
F &=&S(L+X^{-1}R)(E_{p}+Y^{-1/2}O_{p})  \label{propagator} \\
&=&(E_{x}+X^{-1/2}O_{x})(B+Y^{-1}T)S
\end{eqnarray}
\end{definition}

\begin{lemma}
$F$ is unitary.
\end{lemma}

\proof%
Observe that $E_{x}=E_{x}^{\dagger }$, $O_{x}=O_{x}^{\dagger }$, $%
B=B^{\dagger }$, $T=T^{\dagger }$. It follows that 
\begin{eqnarray*}
(L+X^{-1}R)(L+X^{-1}R)^{\dagger } &=&L+R=I, \\
(E_{p}+Y^{-1/2}O_{p})(E_{p}+Y^{-1/2}O_{p})^{\dagger } &=&E_{p}+O_{p}=I.
\end{eqnarray*}
Thus $F$ is the product of three unitary operators, hence is unitary. 
\endproof%

\section{The Classical Limit}

Because of the piecewise continuity of the baker's map (and its covering
map), the classical limit requires more thought. The basic problem comes
from the fact that the projection operators $L$ and $E_{p}$ (for example) do
not commute as $\hbar \rightarrow 0$ (even weakly). This fact comes
basically from scaling - each term contributes less, but the number of terms
increases.

All is not lost, however, as we can take a more constrained view of what
constitutes a quantum state with a classical limit (see, for example,
[Hep]). We take as our quantum state the coherent state $\left| \phi _{\hbar
};x_{0},p_{0}\right\rangle $ centered around the point $\left(
x_{0},p_{0}\right) $ with a width of $\sqrt{\hbar }$. Recall that a coherent
state can be written 
\begin{equation}
\phi _{x_{0},p_{0}}^{\hbar }\left( x\right) =\left\langle x\right. \left|
\phi _{\hbar };x_{0},p_{0}\right\rangle =\frac{1}{\left( \pi \hbar \right)
^{1/4}}e^{-\left( x-x_{0}\right) ^{2}/2\hbar }e^{ip_{0}x/\hbar
-ip_{0}x_{0}/2\hbar }  \label{coherent state}
\end{equation}
with a Fourier transform 
\[
\widetilde{\phi }_{x_{0},p_{0}}^{\hbar }\left( p\right) =\left\langle
p\right. \left| \phi _{\hbar };x_{0},p_{0}\right\rangle =\frac{1}{\left( \pi
\hbar \right) ^{1/4}}e^{-\left( p-p_{0}\right) ^{2}/2\hbar
}e^{-ipx_{0}/\hbar +ip_{0}x_{0}/2\hbar }. 
\]

We are now in a postion to define the classical limit.

\begin{definition}
\label{classlim}We define a quantum propagator $F$ to have a weak classical
limit if for any $A\in \frak{A}_{0}$, and for almost every $x_{0},p_{0},$%
\[
\lim_{\hbar \rightarrow 0}\left( \left\langle \phi _{\hbar
};x_{0},p_{0}\right| F^{\dagger }T_{\hbar }\left( A\right) F\left| \phi
_{\hbar };x_{0},p_{0}\right\rangle -\left\langle \phi _{\hbar };\beta \left(
x_{0},p_{0}\right) \right| T_{\hbar }\left( A\right) \left| \phi _{\hbar
};\beta \left( x_{0},p_{0}\right) \right\rangle \right) =0\text{.} 
\]
\end{definition}

where $\beta $ is the classical evolution.

\smallskip In other words, if, as $\hbar \rightarrow 0$, all observables
have the same values under classical and quantum evolution for almost all
wave packets, we say the quantum mechanics yields the classical mechanics.
Note also that this differs from the definition given in [Hep] by the use of
the ``almost all'' caveat.

With this definition, we have the following theorem.

\begin{theorem}
The propagator $F$ defined in \ref{propagator} has a weak classical limit in
the sense of definition \ref{classlim}.
\end{theorem}

Proof. We give the proof for a harmonic $A=U^{a}V^{b}$. The general case
will follow by linearity and continuity. We divide the proof into steps.

Step 1. We first calculate the expectation value of the operator $U^{a}V^{b}$
in the coherent states. We see 
\begin{eqnarray*}
&&\left\langle \phi _{\hbar };x_{0},p_{0}\right| U^{a}V^{b}\left| \phi
_{\hbar };x_{0},p_{0}\right\rangle \\
&=&\frac{1}{\left( \pi \hbar \right) ^{1/2}}\int dxdx^{\prime }e^{-\left(
x-x_{0}\right) ^{2}/2\hbar }e^{-ip_{0}x/\hbar }e^{2\pi iax}\delta \left(
x-x^{\prime }-2\pi b\hbar \right) e^{-\left( x^{\prime }-x_{0}\right)
^{2}/2\hbar }e^{ip_{0}x^{\prime }/\hbar } \\
&=&e^{2\pi ibp_{0}}e^{2\pi iax_{0}}e^{-2\pi ^{2}b^{2}\hbar }e^{-4\pi
^{2}\hbar \left( b+ia\right) ^{2}}\rightarrow e^{2\pi ibp_{0}}e^{2\pi
iax_{0}}\quad \text{as}\quad \hbar \rightarrow 0\text{.}
\end{eqnarray*}
Step 2. Observe that acting on these states, we see 
\begin{eqnarray*}
\left\| L\left| \phi _{\hbar };x_{0},p_{0}\right\rangle \right\| ^{2}
&=&\left\langle \phi _{\hbar };x_{0},p_{0}\right| L\left| \phi _{\hbar
};x_{0},p_{0}\right\rangle \\
&=&\frac{1}{\left( \pi \hbar \right) ^{1/2}}\sum_{k\in \mathbb{Z}%
}\int_{0}^{1/2}e^{-\left( x+k-x_{0}\right) ^{2}/\hbar }dx.
\end{eqnarray*}
Suppose $x_{0}\neq l/2$ for any $l\in \Bbb{Z}$. Now choose $\epsilon >0$ and 
$l\in \Bbb{Z}$ such that $x_{0}\in [l/2+\epsilon ,\left( l+1\right)
/2-\epsilon )$ for $l\in \mathbb{Z}$ and $\epsilon >0$.

For the case of $l$ odd, we see that the value of the integral is bounded by 
\[
\left| \frac{1}{\left( \pi \hbar \right) ^{1/2}}\sum_{k\in \mathbb{Z}%
}\int_{0}^{1/2}e^{-\left( x-x_{0}+k\right) ^{2}/\hbar }dx\right| \leq \frac{2%
}{\sqrt{\pi }}\int_{\epsilon /\sqrt{\hbar }}^{\infty }e^{-x^{2}}dx. 
\]
A bound on this integral can easily be given for $\epsilon /\sqrt{\hbar }>1.$
We see that 
\[
\int_{\epsilon /\sqrt{\hbar }}^{\infty }e^{-x^{2}}dx\leq \int_{\epsilon /%
\sqrt{\hbar }}^{\infty }xe^{-x^{2}}dx=\frac{1}{2}\int_{\epsilon /\sqrt{\hbar 
}}^{\infty }e^{-u}du=e^{-\epsilon ^{2}/\hbar }/2. 
\]
Thus we see that for $l$ odd, the limit of the integral is zero as $\hbar
\rightarrow 0.$

Now consider $x_{0}\in [l/2+\epsilon ,\left( l+1\right) /2-\epsilon )$ with $%
l$ even. Then we see that 
\begin{eqnarray*}
\left\| L\left| \phi _{\hbar };x_{0},p_{0}\right\rangle \right\| ^{2}
&=&\left\| \left( I-R\right) \left| \phi _{\hbar };x_{0},p_{0}\right\rangle
\right\| ^{2} \\
&=&1-\left\| R\left| \phi _{\hbar };x_{0},p_{0}\right\rangle \right\| ^{2} \\
&\rightarrow &1\quad \text{as\ }\hbar \rightarrow 0\text{.}
\end{eqnarray*}
Similar results hold for all the projection operators $%
L,R,B,T,E_{x},O_{x},E_{p},O_{p}$ defined in equations \ref{Proj1} and \ref
{Proj2}. We let $\tilde{l},\tilde{r},\tilde{b},\tilde{t},\tilde{e}_{x},%
\tilde{o}_{x},\tilde{e}_{p},\tilde{o}_{p}$ denote the interior of the
regions given in \ref{regone} and \ref{regtwo}, that is the regions with the
boundaries removed. (For instance, $\tilde{l}$ does not containt $x=0$ or $%
x=1/2$.) Note that $\left( \tilde{l}\cap \tilde{e}_{p}\right) \cup \left( 
\widetilde{r}\cap \tilde{o}_{p}\right) \cup \left( \widetilde{l}\cap \tilde{o%
}_{p}\right) \cup \left( \widetilde{r}\cap \widetilde{e}_{p}\right) $ is
dense in $\Bbb{R}^{2}$.

We can summarize these results in the following table. 
\[
\begin{array}{ccc}
\text{Operator (}\mathcal{O}\text{)} & \left( x_{0},p_{0}\right) \in \,\,? & 
\lim_{\hbar \rightarrow 0}\left\| \mathcal{O}\left| \phi _{\hbar
};x_{0},p_{0}\right\rangle \right\| \\ 
E_{x} & \tilde{e}_{x} & 1 \\ 
E_{x} & \tilde{o}_{x} & 0 \\ 
O_{x} & \tilde{e}_{x} & 0 \\ 
O_{x} & \tilde{o}_{x} & 1 \\ 
E_{p} & \tilde{e}_{p} & 1 \\ 
E_{p} & \tilde{o}_{p} & 0 \\ 
O_{p} & \tilde{e}_{p} & 0 \\ 
O_{p} & \tilde{o}_{p} & 1 \\ 
L & \widetilde{l} & 1 \\ 
L & \widetilde{r} & 0 \\ 
R & \widetilde{l} & 0 \\ 
R & \widetilde{r} & 1 \\ 
B & \widetilde{b} & 1 \\ 
B & \widetilde{t} & 0 \\ 
T & \widetilde{b} & 0 \\ 
T & \widetilde{t} & 1
\end{array}
\]
Step 3. Now suppose $\left( x_{0},p_{0}\right) \in \tilde{r}\cap \widetilde{o%
}_{p}$. Then consider the quantum evolution. We see 
\begin{eqnarray*}
&&\left\langle \phi _{\hbar };x_{0},p_{0}\right| \left(
E_{p}+Y^{1/2}O_{p}\right) \left( L+XR\right) S^{\dagger }U^{a}V^{b}S\left(
L+X^{-1}R\right) \left( E_{p}+Y^{-1/2}O_{p}\right) \left| \phi _{\hbar
};x_{0},p_{0}\right\rangle \\
&=&\left\langle \phi _{\hbar };x_{0},p_{0}\right| \left(
E_{p}+Y^{1/2}O_{p}\right) \left( L+XR\right) U^{2a}V^{b/2}\left(
L+X^{-1}R\right) \left( E_{p}+Y^{-1/2}O_{p}\right) \left| \phi _{\hbar
};x_{0},p_{0}\right\rangle
\end{eqnarray*}
Multiplying out, we see $16$ term in the expansion. Consider one of these
terms. We see from the chart that 
\begin{eqnarray*}
&&\left\| \left\langle \phi _{\hbar };x_{0},p_{0}\right|
E_{p}XRU^{2a}V^{b/2}LY^{-1/2}O_{p}\left| \phi _{\hbar
};x_{0},p_{0}\right\rangle \right\| \\
&=&\left\| \left\langle \phi _{\hbar };x_{0},p_{0}\right|
E_{p}RU^{2a+N}V^{\left( b-N\right) /2}RO_{p}\left| \phi _{\hbar
};x_{0},p_{0}\right\rangle \right\| \\
&\leq &\left\| \left\langle \phi _{\hbar };x_{0},p_{0}\right| E_{p}\left|
\phi _{\hbar };x_{0},p_{0}\right\rangle \right\| \\
&&\times \left\| \left\langle \phi _{\hbar };x_{0},p_{0}\right| \left(
RU^{2a+N}V^{\left( b-N\right) /2}RO_{p}\right) ^{\dagger }\left(
RU^{2a+N}V^{\left( b-N\right) /2}RO_{p}\right) \left| \phi _{\hbar
};x_{0},p_{0}\right\rangle \right\| \\
&\rightarrow &0\text{\quad as\quad }\hbar \rightarrow 0\text{.}
\end{eqnarray*}

Similarly, $15$ of the terms vanish as $\hbar \rightarrow 0$. The only
surviving term for $\left( x_{0},p_{0}\right) \in \tilde{r}\cap \widetilde{o}%
_{p}$ is 
\begin{eqnarray*}
&&\left\langle \phi _{\hbar };x_{0},p_{0}\right|
O_{p}Y^{1/2}LU^{2a}V^{b/2}LY^{-1/2}O_{p}\left| \phi _{\hbar
};x_{0},p_{0}\right\rangle \\
&\rightarrow &\left\langle \phi _{\hbar };x_{0},p_{0}\right|
U^{2a}V^{b/2}\left| \phi _{\hbar };x_{0},p_{0}\right\rangle \quad \\
&\rightarrow &e^{2\pi i\left( 2ax_{0}+\left( b/2\right) p_{0}\right) }.
\end{eqnarray*}
Step 4. Now consider the classical evolution. We have, for $\left(
x_{0},p_{0}\right) \in \tilde{r}\cap \widetilde{o}_{p}$%
\begin{eqnarray*}
\left\langle \phi _{\hbar };\beta \left( x_{0},p_{0}\right) \right|
U^{a}V^{b}\left| \phi _{\hbar };\beta \left( x_{0},p_{0}\right)
\right\rangle &=&\left\langle \phi _{\hbar };2x_{0},p_{0}/2\right|
U^{a}V^{b}\left| \phi _{\hbar };2x_{0},p_{0}/2\right\rangle \\
&\rightarrow &e^{2\pi i\left( 2ax_{0}+\left( b/2\right) p_{0}\right) }\text{.%
}
\end{eqnarray*}
The calculations for the other three regions are similar, and we omit the
details here.This concludes the proof. 
\endproof%

\section{\protect\smallskip Planck's Constant $=1/N$}

A remarkable set of properties can be associated with quantum dynamics on a
torus if we let Planck's constant satisfy the integrality condition 
\[
h=1/N. 
\]
This fact is evidenced by the quantization schemes presented in [BV] and [S]
for the baker's map, and [BV] for the cat maps. In [KLMR] and [LRS], an
explicit construction similar to what we presented in the previous section
was given for the cat map, kick maps, and Harper maps. The Hilbert space is
taken to be the standard $L^{2}\left( \Bbb{R}\right) $. The quantum torus is
defined as the algebra of observables, or operators on $L^{2}\left( \Bbb{R}%
\right) $, generated by $\left\{ \exp \left( 2\pi i\widehat{x}\right) ,\exp
\left( 2\pi i\widehat{p}\right) \right\} $. In [KLMR] a propagator is found
which yields the classical dynamics as $\hbar \rightarrow 0$ for the cat
map. This quantization scheme was called ``the quantum cat map.'' It is
valid for all $\hbar $, but was shown to reduce to the quantizations given
in [BV] for $h=1/N.$ Here we present a similar result for the baker's map.
We shall find for the finite-dimensional matrix propagator for $N$ even 
\begin{eqnarray*}
&&\left( \Phi _{n}^{\left( 0,0\right) },F\Phi _{m}^{\left( 0,0\right)
}\right) _{P} \\
&=&\left\{ 
\begin{array}{ll}
\sum_{a=0}^{N/2-1}\left( \mathcal{F}^{N}\right) _{na}^{-1}\left( 
\begin{array}{ll}
\mathcal{F}^{N/2} & 0 \\ 
0 & \mathcal{F}^{N/2}
\end{array}
\right) _{am} & n\text{ even} \\ 
\sum_{a=0}^{N/2-1}\left( \mathcal{F}^{N}\right) _{na}^{-1}\left( 
\begin{array}{ll}
e^{i\pi \left( n-2m\right) /N}\mathcal{F}^{N/2} & 0 \\ 
0 & e^{i\pi \left( n-\left( 2m-N\right) \right) /N}\mathcal{F}^{N/2}
\end{array}
\right) _{am} & n\,\text{odd.}
\end{array}
\right.
\end{eqnarray*}
where $\mathcal{F}^{N}$ is the matrix for the $N$-dimensional discrete
Fourier transform, and $\Phi _{n}^{\left( 0,0\right) }$ is a basis vector of
the Hilbert space $\mathcal{H}_{\hbar }\left( 0\right) \cong \Bbb{C}^{N}$
defined as the periodic $\delta $-comb \ref{delta comb}, and $\left( \cdot
,\cdot \right) $ is the inner product defined in equation \ref{ip}.

This propagator is shown here to preserve the symmetry $x\rightarrow 1-x$
and $p\rightarrow 1-p$, which is not preserved in the original quantization
given by Balazs and Voros [BV]. (In Saraceno [S], an anti-periodic
quantization is formulated which does preserve this symmetry.) Here we see
that with a small ($O\left( \hbar \right) $) correction to the Balazs-Voros
matrices, the periodic quantization can also be made to preserve the
classical symmetries. Note that along the classical trajectories $n=2m$ or $%
n=2m-N$, the extra phase we obtain using this quantization vanishes.

\section{\protect\smallskip The $\theta $-torus}

We let $U=\exp \left( 2\pi i\widehat{x}\right) $ and $V=\exp \left( 2\pi i%
\widehat{p}\right) $ be operators on an infinite dimensional Hilbert space
(Bargmann space $\mathcal{H}^{2}\left( \mathbb{C},d\mu _{\hbar }\right) $),
with $\left[ \widehat{x},\widehat{p}\right] =i\hbar $. Observe that for $%
h=1/N$ the algebra has a natural center generated by 
\begin{eqnarray}
X &=&U^{N},  \label{center} \\
Y &=&V^{N}.  \nonumber
\end{eqnarray}
That is 
\[
\left[ X,Y\right] =\left[ X,U\right] =\left[ X,V\right] =\left[ Y,U\right]
=\left[ Y,V\right] =0. 
\]
In [KLMR], this insight was used to show that Bargmann space (the Hilbert
space of entire functions on the plane) can be decomposed via the following
eigenvalue problem: 
\begin{eqnarray*}
X\phi \left( z\right) &=&e^{2\pi i\theta _{1}}\phi \left( z\right) , \\
Y\phi \left( z\right) &=&e^{2\pi i\theta _{2}}\phi \left( z\right) ,
\end{eqnarray*}
where $\theta =\left( \theta _{1},\theta _{2}\right) \in \Bbb{T}^{2}$. As in
[KLMR], let $\mathcal{H}_{\hbar }\left( \theta \right) $ denote the space of
(non-normalizable) independent eigenvectors with fixed $\theta $. The space $%
\mathcal{H}_{\hbar }\left( \theta \right) $ was shown to have a natural
inner product defined as an integral over the fundamental domain $%
D=[0,1]\times [0,1]\subset \Bbb{C}$ of Bargmann space given by 
\begin{equation}
\left( \phi _{1},\phi _{2}\right) _{P}=\int_{D}\overline{\phi _{1}\left(
z\right) }\phi _{2}\left( z\right) d\mu _{\hbar }\left( z\right) \text{,}
\label{inner product}
\end{equation}
where $d\mu _{\hbar }\left( z\right) =\left( \pi \hbar \right) ^{-1}\exp
\left( -\left| z\right| ^{2}/\hbar \right) d^{2}z$. An explicit isomorphism $%
\kappa :\mathcal{H}^{2}\left( \mathbb{C},d\mu _{\hbar }\right) \rightarrow
\int_{T^{2}}^{\oplus }\mathcal{H}_{\hbar }\left( \theta \right) d\theta $
was also derived between $\mathcal{H}^{2}\left( \mathbb{C},d\mu _{\hbar
}\right) $ and the direct integral of the spaces $\mathcal{H}_{\hbar }\left(
\theta \right) $. The isomorphism was shown to be inner product preserving, 
\[
\left( \psi _{1},\psi _{2}\right) _{\mathcal{\mathcal{H}}^{2}}=\int_{%
\mathbb{T}^{2}}\left( \kappa \psi _{1}\left( \theta \right) ,\kappa \psi
_{2}\left( \theta \right) \right) _{P}d^{2}\theta , 
\]
where $\psi _{i}=\int_{\Bbb{T}^{2}}\kappa \psi _{i}\left( \theta \right)
d\theta $. The following lemma was proved:

\begin{lemma}
\label{bargman theta}(i) The following functions are elements of $\mathcal{H}%
_{\hbar }\left( \theta \right) $ of unit norm: 
\begin{equation}
\phi _{m}^{\left( \theta \right) }\left( z\right) =C_{m}\left( \theta
\right) e^{-N\pi z^{2}+2\sqrt{2}\pi \left( \theta _{1}+m\right) z}\sum_{k\in %
\mathbb{Z}}e^{-N\pi k^{2}-2\pi \left( \theta _{1}+i\theta _{2}+m\right) k+2%
\sqrt{2}N\pi kz},  \label{basis functions}
\end{equation}
where 
\[
C_{m}\left( \theta \right) :=\left( 2/N\right) ^{1/4}e^{-\pi \left( \theta
_{1}+m\right) ^{2}/N-2\pi i\theta _{2}m/N}. 
\]
They are periodic in $m$, 
\begin{equation}
\phi _{m+N}^{\left( \theta \right) }=\phi _{m}^{\left( \theta \right) },
\label{periodicity}
\end{equation}
and furthermore, 
\[
\phi _{0}^{\left( \theta \right) },...,\phi _{N-1}^{\left( \theta \right) } 
\]
are orthogonal vectors in $\mathcal{H}_{\hbar }\left( \theta \right) $.

(ii) The space $\mathcal{H}_{\hbar }\left( \theta \right) $ has dimension $N$%
. Consequently the functions \ref{basis functions} form an orthonormal basis
for $\mathcal{H}_{\hbar }\left( \theta \right) $.
\end{lemma}

Composing the Bargmann transformation $\mathcal{H}^{2}\left( \mathbb{C},d\mu
_{\hbar }\right) $ $\rightarrow $ $L^{2}\left( \mathbb{R},dx\right) $ with
this isomorphism, we can construct the transformation between $L^{2}\left( %
\mathbb{R},dx\right) $ and $\int_{T^{2}}^{\oplus }\mathcal{H}_{\hbar }\left(
\theta \right) d\theta $. Applying the Bargmann transformation to the basis
functions $\phi _{m}^{(\theta )}\in \mathcal{H}_{\hbar }(\theta )$, we find 
\begin{equation}
\Phi _{m}^{(\theta )}(x)=B^{-1}\phi _{m}^{(\theta )}(x)=\frac{e^{2\pi
i\theta _{2}m/N}}{N^{1/2}}\sum_{k\in \mathbb{Z}}e^{2\pi i\theta _{2}k}\delta
\left( x-\frac{m+\theta _{1}+NK}{N}\right) .  \label{delta comb}
\end{equation}
This is the $\delta $-comb wavefunctions described informally in the physics
literature (see, for example [HB]).We use the following notation for these
vectors: 
\[
\Phi _{m}^{(\theta )}=\frac{e^{2\pi i\theta _{2}m/N}}{N^{1/2}}\sum_{k\in %
\mathbb{Z}}e^{2\pi i\theta _{2}k}\left| \frac{\theta _{1}+m}{N}%
+k\right\rangle _{x} 
\]

We can, of course, just as easily work in momentum representation. In fact,
for $h=1/N$, a rather interesting calculational identity can be found.

\begin{lemma}
For $h=1/N$, 
\[
\Phi _{m}^{(\theta )}=e^{-2\pi i\theta _{1}\theta _{2}/N}\sum_{n=0}^{N-1}%
\mathcal{F}_{mn}^{N}\widetilde{\Phi }_{n}^{(\theta )}, 
\]
where $\left\{ \widetilde{\Phi }_{n}^{(\theta )}\right\} _{0\leq n\leq N-1}$
are the momentum-state wave functions on the torus, 
\begin{equation}
\widetilde{\Phi }_{n}^{(\theta )}=\frac{e^{-2\pi in\theta _{1}/N}}{\sqrt{N}}%
\sum_{k}e^{-2\pi i\theta _{1}k}\left| \frac{\theta _{2}+n}{N}+k\right\rangle
_{p},  \label{p state delta comb}
\end{equation}
and $\mathcal{F}_{mn}^{N}$ is the matrix for the discrete Fourier transform, 
\[
\mathcal{F}_{mn}^{N}=\frac{e^{-2\pi imn/N}}{\sqrt{N}}. 
\]
\end{lemma}

\begin{remark}
We see in particular that for the subsets $\theta _{1}=0$ or $\theta _{2}=0$%
, changing coordinates from momentum representation to position
representation is simply a discrete Fourier transform.
\end{remark}

\proof%
The proof is a direct calculation. We have 
\begin{eqnarray*}
\int_{\mathbb{R}}\left| p\right\rangle \left\langle p\right| \Phi
_{m}^{(\theta )}dp &=&\frac{e^{2\pi i\theta _{2}m/N}}{N^{1/2}}\sum_{k\in %
\mathbb{Z}}e^{2\pi i\theta _{2}k}\int_{\mathbb{R}}\left| p\right\rangle
\left\langle p{\Huge |}\frac{m+\theta _{1}}{N}+k\right\rangle _{x}dp \\
&=&\frac{e^{2\pi i\theta _{2}m/N}}{N}\sum_{k\in \mathbb{Z}}\left| \frac{%
\theta _{2}+k}{N}\right\rangle _{p}\exp \left\{ -2\pi i\left( \frac{\theta
_{2}+k}{N}\right) (m+\theta _{1})\right\} .
\end{eqnarray*}
We now let $k\rightarrow n+kN$, with $n\in \left\{ 0,...,N-1\right\} $, and $%
k\in \mathbb{Z}$, to find 
\begin{eqnarray*}
\int_{\mathbb{R}}\left| p\right\rangle \left\langle p\right| \Phi
_{m}^{(\theta )}dp &=&\frac{e^{-2\pi i\theta _{1}\theta _{2}/N}}{N}%
\sum_{n}e^{-2\pi imn/N}e^{-2\pi in\theta _{1}/N}\sum_{k}e^{-2\pi i\theta
_{1}k}\left| \frac{\theta _{2}+n}{N}+k\right\rangle _{p} \\
&=&e^{-2\pi i\theta _{1}\theta _{2}/N}\sum_{n}\mathcal{F}_{mn}\widetilde{%
\Phi }_{n}^{(\theta )},
\end{eqnarray*}
as claimed. 
\endproof%

Analogous to (\ref{inner product}), we can also find an explicit expression
for the inner product over the $N$-dimensional Hilbert space at each point
on the $\theta $-torus as an integral over the fundamental domain $[0,1]$ of
the real line.

The inner product defined in \ref{inner product} can be written as 
\begin{equation}
\left( \Psi _{1}(\theta ),\Psi _{2}(\theta )\right) _{P}=\int_{0}^{1}%
\overline{\Psi _{1}(x,\theta )}(K\Psi _{2})(x,\theta )dx,  \label{ip}
\end{equation}
where 
\begin{eqnarray*}
K\Psi _{2}(x,\theta ) &=&\int_{-\infty }^{\infty }K\left( x,y\right) \Psi
_{2}(y,\theta )dy, \\
K\left( x,y\right) &=&\frac{1}{2\pi \hbar }g\left( \frac{x-y}{2\hbar }\right)
\end{eqnarray*}
and 
\[
g(r)=\frac{\sin r}{r}\,e^{-\hbar r^{2}+ir}. 
\]

We see this via a direct calculation: 
\begin{eqnarray*}
\int_{D}\overline{\psi _{1}(z,\theta )}\psi _{2}(z,\theta )d\mu _{\hbar }(z)
&=&\int_{D}\overline{B\Psi _{1}(z,\theta )}\Psi _{2}(z,\theta )d\mu _{\hbar
}(z) \\
&=&\frac{1}{2\left( \pi \hbar \right) ^{3/2}}\sum_{k\in \mathbb{Z}%
}\int_{0}^{1}dx\int_{-\infty }^{\infty }dy\overline{\Psi _{1}(x+k,\theta )}%
\Psi _{2}(y,\theta )e^{-((x+k)^{2}+y^{2})/2\hbar } \\
&&\times \int_{D}e^{-(u^{2}-(x+k)(u-iv)+y(u+iv))/\hbar }dudv.
\end{eqnarray*}
We next use the fact that both $\Psi _{1}$ and $\Psi _{2}$ satisfy $X\Psi
_{i}=e^{2\pi i\theta _{1}}\Psi _{i}$ and $Y\Psi _{i}=e^{2\pi i\theta
_{2}}\Psi _{i}$. Substituting in, we find 
\begin{eqnarray*}
&&\frac{1}{2\left( \pi \hbar \right) ^{3/2}}\sum_{k\in \mathbb{Z}%
}\int_{0}^{1}dx\int_{-\infty }^{\infty }dye^{-2\pi ik\theta _{2}}\overline{%
\Psi _{1}(x,\theta )}\Psi _{2}(y,\theta )e^{-((x+k)^{2}+y^{2})/2\hbar } \\
&&\times \int_{D}dudve^{-(u^{2}-(x+k)(u-iv)+y(u+iv))/\hbar } \\
&=&\frac{1}{2\left( \pi \hbar \right) ^{3/2}}\sum_{k\in \mathbb{Z}%
}\int_{0}^{1}dx\overline{\Psi _{1}(x,\theta )}\sum_{k\in \mathbb{Z}%
}\int_{-\infty }^{\infty }dy\Psi _{2}(y-k,\theta
)e^{-((x+k)^{2}+y^{2})/2\hbar } \\
&&\times \int_{D}e^{-(u^{2}-(x+k)(u-iv)+y(u+iv))/\hbar }dudv \\
&=&\frac{1}{2\left( \pi \hbar \right) ^{3/2}}\int_{0}^{1}dx\overline{\Psi
_{1}(x,\theta )}\int_{-\infty }^{\infty }dy\Psi _{2}(y,\theta )\frac{\sin
\left( \frac{x-y}{2\hbar }\right) }{\left( \frac{x-y}{2\hbar }\right) }%
e^{-(x^{2}+y^{2}+i(x-y))/2\hbar } \\
&&\times \int_{-\infty }^{\infty }e^{-((u-k)^{2}-(x+y)(u-k))/\hbar }du \\
&=&\int_{0}^{1}\overline{\Psi _{1}(x,\theta )}(K\Psi _{2})(x,\theta )dx.
\end{eqnarray*}

We see that the kernel $K(x,y)$ is a type of quantum diffraction in keeping
with the uncertainty principle. In fact it can be shown that as $\hbar
\rightarrow 0$, $K(x,y)\rightarrow \delta (x-y)$. Observe also that with
respect to this inner product, the basis elements $\left\{ \Phi
_{m}^{(\theta )}\right\} $ are orthonormal: $\left( \Phi _{m}^{(\theta
)},\Phi _{n}^{(\theta )}\right) _{P}=\delta _{mn}.$

\section{Dynamics at $\theta =\left( 0,0\right) $}

The point $\theta =\left( 0,0\right) $ of the $\theta $-torus corresponds to
the $N$-dimensional vector space $\mathcal{H}_{\hbar }\left( 0\right) $ of
periodic $\delta $-combs. In fact, for the quantum baker's map, we now show
that $\theta =$ $\left( 0,0\right) $ is a fixed point of the dynamics on the 
$\theta $-torus \textit{for }$N$\textit{\ even}. That is, the set of
periodic $\delta $-combs is mapped onto itself by our propagator $F$. We
have 
\[
XF\Phi _{m}^{\left( 0,0\right) }=FX^{2}\Phi _{m}^{\left( 0,0\right) }=F\Phi
_{m}^{\left( 0,0\right) } 
\]
and also 
\begin{eqnarray*}
YF\phi _{m}^{\left( 0,0\right) } &=&\left( O_{x}+X^{-1/2}E_{x}\right) \left(
B+Y^{-1}T\right) SY^{1/2}\Phi _{m}^{\left( 0,0\right) } \\
&=&\left( E_{x}+X^{1/2}O_{x}\right) \left( B+YT\right) SXY\Phi _{m}^{\left(
0,0\right) } \\
&=&\left( E_{x}+X^{-1/2}O_{x}\right) \left( B+Y^{-1}T\right) SXY\Phi
_{m}^{\left( 0,0\right) } \\
&=&F\Phi _{m}^{\left( 0,0\right) }.
\end{eqnarray*}
Furthermore, at $\theta =\left( 0,0\right) $ the observables corresponding
to the algebra $\frak{A}_{\hbar }$ generated by $\exp \left( 2\pi i\widehat{x%
}\right) $ and $\exp \left( 2\pi i\widehat{p}\right) $characterized by the
equations $\left[ X,A\right] =\left[ Y,A\right] =0$ for $A\in \frak{A}%
_{\hbar }$ is also preserved by the quantum dynamics.That is, we have the
following lemma.

\begin{lemma}
For $N$ even, and $A\in \frak{A}_{\hbar }$. 
\[
\left[ X,F^{\dagger }AF\right] \Phi _{m}^{\left( 0,0\right) }=\left[
Y,F^{\dagger }AF\right] \Phi _{m}^{\left( 0,0\right) }=0. 
\]
.
\end{lemma}

\begin{remark}
Observe that we can write any $A\ $as $\sum_{j,k}\gamma _{jk}U^{j}V^{k}$.
Letting $\,j=a+Nc$ and $k=b+Nd$ with $0\leq a,b\leq N-1$ and $c,d\in \Bbb{Z}$%
, we see from equation \ref{center} that 
\[
U^{a+Nc}V^{b+Nd}\Phi _{m}^{\left( 0,0\right) }=U^{a}V^{b}X^{c}Y^{d}\Phi
_{m}^{\left( 0,0\right) }=U^{a}V^{b}\Phi _{m}^{\left( 0,0\right) }\text{.} 
\]
Thus, acting on the subspace $\mathcal{H}\left( 0\right) $, the algebra $%
\frak{A}_{\hbar }$ is reduced to a set of $N^{2}$ operators. This is
isomorphic to the algebra of $N\times N$ matrices.
\end{remark}

Proof. The proof is a straightforward calculation using the commutation
relations \ref{commutation relations}. We provide here the case of a pure
harmonic $U^{m}V^{n}$. The general case follows immediately by linearity and
continuity.

\begin{eqnarray*}
&&XF^{\dagger }AF\Phi _{m}^{\left( 0,0\right) } \\
&=&XS^{\dagger }\left( B+YT\right) \left( E_{x}+X^{1/2}O_{x}\right)
U^{m}V^{n}\left( E_{x}+X^{-1/2}O_{x}\right) \left( B+Y^{-1}T\right) S\Phi
_{m}^{\left( 0,0\right) } \\
&=&S^{\dagger }\left( -1\right) ^{n}\left( T+YB\right) \left(
E_{x}+X^{1/2}O_{x}\right) U^{m}V^{n}\left( E_{x}+X^{-1/2}O_{x}\right) \left(
T+Y^{-1}B\right) SX\Phi _{m}^{\left( 0,0\right) } \\
&=&S^{\dagger }\left( B+YT\right) \left( E_{x}+X^{1/2}O_{x}\right)
U^{m}V^{n}\left( E_{x}+X^{-1/2}O_{x}\right) \left( B+Y^{-1}T\right) S\Phi
_{m}^{\left( 0,0\right) } \\
&=&F^{\dagger }AF\Phi _{m}^{\left( 0,0\right) }\,.
\end{eqnarray*}
Likewise, 
\[
YF^{\dagger }AF\Phi _{m}^{\left( 0,0\right) }=F^{\dagger }AF\Phi
_{m}^{\left( 0,0\right) }. 
\]
\endproof%

We can now determine the matrix elements for the dynamics at this fixed
point. This result should be compared to the baker's map quantum propagator
given in [BV].

\begin{theorem}
The matrix elements for the propagator $F$ on the subspace $\mathcal{H}%
\left( 0\right) $ are given by 
\begin{eqnarray}
&&\left( \Phi _{n}^{(0,0)},F\Phi _{m}^{(0,0)}\right) _{P}  \label{matrix} \\
&=&\left\{ 
\begin{array}{ll}
\sum_{a=0}^{N-1}\left( \mathcal{F}^{N}\right) _{na}^{-1}\left( 
\begin{array}{ll}
\mathcal{F}^{N/2} & 0 \\ 
0 & \mathcal{F}^{N/2}
\end{array}
\right) _{am} & n\text{ even} \\ 
\sum_{a=0}^{N-1}\left( \mathcal{F}^{N}\right) _{na}^{-1}\left( 
\begin{array}{ll}
e^{i\pi \left( n-2m\right) /N}\mathcal{F}^{N/2} & 0 \\ 
0 & e^{i\pi \left( n-\left( 2m-N\right) \right) /N}\mathcal{F}^{N/2}
\end{array}
\right) _{am} & n\,\text{odd}
\end{array}
\right.
\end{eqnarray}
\end{theorem}

Proof. We divide this calculation into different cases. For $0\leq m<N/2$,
we have 
\begin{eqnarray*}
F\Phi _{m}^{\left( 0,0\right) } &=&\left( B+Y^{-1}T\right) \left(
E_{x}+X^{-1/2}O_{x}\right) S\Phi _{m}^{\left( 0,0\right) } \\
&=&\left( B+Y^{-1}T\right) \left( E_{x}+X^{-1/2}O_{x}\right) S\frac{1}{\sqrt{%
N}}\sum_{k\in \mathbb{Z}}\left| \frac{m}{N}+k\right\rangle \\
&=&\left( B+Y^{-1}T\right) \left( E_{x}+X^{-1/2}O_{x}\right) \sqrt{\frac{2}{N%
}}\sum_{k\in \mathbb{Z}}\left| \frac{2m}{N}+2k\right\rangle \\
&=&\frac{1}{\sqrt{2}}\left( B+Y^{-1}T\right) \left(
E_{x}+X^{-1/2}O_{x}\right) \left( \Phi _{2m}^{\left( 0,0\right) }+e^{-2\pi
im/N}\Phi _{2m}^{\left( 0,1/2\right) }\right) \\
&=&\frac{1}{\sqrt{2}}\Phi _{2m}^{\left( 0,0\right) }+\frac{e^{-2\pi im/N}}{%
\sqrt{2}}\left( B+Y^{-1}T\right) \left( E_{x}+X^{-1/2}O_{x}\right) \Phi
_{2m}^{\left( 0,1/2\right) }.
\end{eqnarray*}
\begin{eqnarray*}
F\Phi _{m}^{\left( 0,0\right) } &=&\left( E_{x}+X^{-1/2}O_{x}\right) \left(
B+Y^{-1}T\right) S\Phi _{m}^{\left( 0,0\right) } \\
&=&\left( E_{x}+X^{-1/2}O_{x}\right) \left( B+Y^{-1}T\right) \sqrt{\frac{2}{N%
}}\sum_{k\in \mathbb{Z}}\left| \frac{2m}{N}+2k\right\rangle \\
&=&\frac{1}{\sqrt{2}}\left( E_{x}+X^{-1/2}O_{x}\right) \left(
B+Y^{-1}T\right) \left( \Phi _{2m}^{\left( 0,0\right) }+e^{-2\pi im/N}\Phi
_{2m}^{\left( 0,1/2\right) }\right) \\
&=&\frac{1}{\sqrt{2}}\Phi _{2m}^{\left( 0,0\right) }+\frac{e^{-2\pi im/N}}{%
\sqrt{2}}\left( E_{x}+X^{-1/2}O_{x}\right) \left( B-T\right) \Phi
_{2m}^{\left( 0,1/2\right) }.
\end{eqnarray*}
Now, observe that 
\begin{eqnarray*}
B\Phi _{m}^{(0,1/2)} &=&\frac{e^{i\pi m/N}}{\sqrt{N}}\sum_{k\in \mathbb{Z}%
}(-1)^{k}\int_{[0,1/2)+\mathbb{Z}}\left| p\right\rangle \left\langle
p\right| \left. \frac{m}{N}+k\right\rangle _{x}dp \\
&=&e^{i\pi m/N}\int_{[0,1/2)+\mathbb{Z}}\left| p\right\rangle \left(
\sum_{k\in \mathbb{Z}}e^{2\pi ik(1/2-Np)}\right) e^{-2\pi ipm}dp \\
&=&\frac{e^{i\pi m/N}}{\sqrt{N}}\sum_{a=0}^{N/2-1}e^{-2\pi i\left(
a+1/2\right) m/N}\;\frac{1}{\sqrt{N}}\sum_{k\in \mathbb{Z}}\left| \frac{a+1/2%
}{N}+k\right\rangle _{p} \\
&=&\frac{1}{\sqrt{N}}\sum_{a=0}^{N/2-1}e^{-2\pi iam/N}\;\widetilde{\Phi }%
_{a}^{(0,1/2)},
\end{eqnarray*}
where we have used the ``p-state'' $\delta $-comb given in \ref{p state
delta comb}. Thus, we see that 
\[
\left( B-T\right) \Phi _{m}^{(0,1/2)}=\frac{1}{\sqrt{N}}\left(
\sum_{a=0}^{N/2-1}e^{-2\pi iam/N}-\sum_{a=N/2}^{N-1}e^{-2\pi iam/N}\right)
\sum_{b=0}^{N-1}\left( \mathcal{F}^{-1}\right) _{ab}\;\Phi _{b}^{(0,1/2)}, 
\]
and 
\begin{eqnarray*}
F\Phi _{m}^{\left( 0,0\right) } &=&\frac{1}{\sqrt{2}}\Phi _{2m}^{\left(
0,0\right) } \\
&&+\frac{e^{-2\pi im/N}}{\sqrt{2N}}\left( \sum_{a=0}^{N/2-1}e^{-2\pi
ia\left( 2m\right) /N}-\sum_{a=N/2}^{N-1}e^{-2\pi ia\left( 2m\right)
/N}\right) \\
&&\times \sum_{b=0}^{N-1}\left( \mathcal{F}^{-1}\right) _{ab}\;\left(
E_{x}+\left( -1\right) ^{b}O_{x}\right) \Phi _{b}^{(0,1/2)}.
\end{eqnarray*}
Now, observe 
\[
E_{x}\Phi _{m}^{(0,0)}=\frac{1}{\sqrt{N}}\sum_{k\in \mathbb{Z}}\chi _{e_{x}}(%
\frac{m}{N}+k)\left| \frac{m}{N}+k\right\rangle _{x} 
\]
So for $m\in \left[ 0,N-1\right] +2N\mathbb{Z}$, this yields 
\begin{eqnarray*}
E_{x}\Phi _{m}^{(0,0)} &=&\frac{1}{\sqrt{N}}\sum_{k\in \mathbf{even}}\left| 
\frac{m}{N}+k\right\rangle _{x}=\frac{1}{\sqrt{N}}\sum_{k\in \mathbb{Z}%
}\left| \frac{m}{N}+2k\right\rangle _{x} \\
&=&\frac{\Phi _{m}^{(0,0)}+e^{-i\pi m/N}\Phi _{m}^{(0,1/2)}}{2}.
\end{eqnarray*}
For $m\in \left[ N,2N-1\right] +2N\mathbb{Z}$, however, we find 
\[
E_{x}\Phi _{m}^{(0,0)}=\frac{\Phi _{m}^{(0,0)}-e^{-i\pi m/N}\Phi
_{m}^{(0,1/2)}}{2}. 
\]
We let $\left[ m/N\right] $ be the integer part of $m/N$, and observe that 
\begin{eqnarray*}
E_{x}\Phi _{m}^{(0,0)} &=&\frac{1}{2}\left( \Phi _{m}^{(0,0)}+e^{-i\pi
\left( m/N-\left[ m/N\right] \right) }\Phi _{m}^{(0,1/2)}\right) , \\
O_{x}\Phi _{m}^{(0,0)} &=&\frac{1}{2}\left( \Phi _{m}^{(0,0)}-e^{-i\pi
\left( m/N-\left[ m/N\right] \right) }\Phi _{m}^{(0,1/2)}\right) ,
\end{eqnarray*}
and 
\[
\left( E_{x}-O_{x}\right) \Phi _{m}^{(0,0)}=e^{-i\pi \left( m/N-\left[
m/N\right] \right) }\Phi _{m}^{(0,1/2)}. 
\]
Having checked that this is consistent with $\Phi _{m+N}^{(\theta )}=\Phi
_{m}^{(\theta )}$, we now restrict ourselves to the original basis vectors, $%
m\in [0,N-1]$.

From the identity $\left( E_{x}-O_{x}\right) ^{2}=I$, it follows immediately
that 
\[
\left( E_{x}-O_{x}\right) \Phi _{m}^{(0,1/2)}=e^{i\pi m/N}\Phi _{m}^{(0,0)}. 
\]
Thus, 
\begin{eqnarray*}
F\Phi _{m}^{\left( 0,0\right) } &=&\frac{1}{\sqrt{2}}\Phi _{2m}^{\left(
0,0\right) } \\
&&+\frac{e^{-2\pi im/N}}{\sqrt{2N}}\left( \sum_{a=0}^{N/2-1}e^{-2\pi
ia\left( 2m\right) /N}-\sum_{a=N/2}^{N-1}e^{-2\pi ia\left( 2m\right)
/N}\right) \sum_{b\,\text{even}}^{N-1}\left( \mathcal{F}^{-1}\right)
_{ab}\Phi _{b}^{(0,1/2)} \\
&&+\frac{e^{-2\pi im/N}}{\sqrt{2N}}\left( \sum_{a=0}^{N/2-1}e^{-2\pi
ia\left( 2m\right) /N}-\sum_{a=N/2}^{N-1}e^{-2\pi ia\left( 2m\right)
/N}\right) \sum_{b\,\text{odd}}^{N-1}e^{i\pi b/N}\left( \mathcal{F}%
^{-1}\right) _{ab}\;\Phi _{b}^{(0,0)}.
\end{eqnarray*}
Consider just the middle term. Since we have already shown that $\theta
=\left( 0,0\right) $ is a fixed point of the dynamics, we should see this
term exactly vanishing. In fact a direct calculation readily shows this. For 
$b$ even 
\begin{eqnarray*}
&&\frac{e^{-2\pi im/N}}{\sqrt{2N}}\left( \sum_{a=0}^{N/2-1}e^{-2\pi ia\left(
2m\right) /N}-\sum_{a=N/2}^{N-1}e^{-2\pi ia\left( 2m\right) /N}\right)
\left( \mathcal{F}^{-1}\right) _{ab}\; \\
&=&\frac{e^{-2\pi im/N}}{N\sqrt{2}}\left( \sum_{a=0}^{N/2-1}e^{-2\pi
ia\left( 2m\right) /N}e^{2\pi iab/N}-\sum_{a=0}^{N/2-1}e^{-2\pi i\left(
a+N/2\right) \left( 2m\right) /N}e^{2\pi i\left( a+N/2\right) b/N}\right) \\
&=&0.
\end{eqnarray*}
Thus, 
\begin{eqnarray*}
F\Phi _{m}^{\left( 0,0\right) } &=&\frac{1}{\sqrt{2}}\Phi _{2m}^{\left(
0,0\right) } \\
&&+\frac{e^{-2\pi im/N}}{\sqrt{2N}}\left( \sum_{a=0}^{N/2-1}e^{-2\pi
ia\left( 2m\right) /N}-\sum_{a=N/2}^{N-1}e^{-2\pi ia\left( 2m\right)
/N}\right) \sum_{b\,\text{odd}}^{N-1}e^{i\pi b/N}\left( \mathcal{F}%
^{-1}\right) _{ab}\;\Phi _{b}^{(0,0)}.
\end{eqnarray*}
We next calculate the matrix elements. We see 
\[
\left( \Phi _{n}^{\left( 0,0\right) },F\Phi _{m}^{\left( 0,0\right) }\right)
_{P}=\left\{ 
\begin{array}{ll}
\sum_{a=0}^{N/2-1}\left( \mathcal{F}^{N}\right) _{na}^{-1}\left( 
\begin{array}{ll}
\mathcal{F}^{N/2} & 0 \\ 
0 & 0
\end{array}
\right) _{am} & n\text{ even} \\ 
e^{i\pi \left( n-2m\right) /N}\sum_{a=0}^{N/2-1}\left( \mathcal{F}%
^{N}\right) _{na}^{-1}\left( 
\begin{array}{ll}
\mathcal{F}^{N/2} & 0 \\ 
0 & 0
\end{array}
\right) _{am} & n\,\text{odd}
\end{array}
\right. 
\]
where 
\[
\mathcal{F}_{mn}=\frac{e^{-2\pi imn/N}}{\sqrt{N}}. 
\]
For the case $N/2\leq m<N$, we see 
\begin{eqnarray*}
F\Phi _{m}^{\left( 0,0\right) } &=&\left( E_{x}+X^{-1/2}O_{x}\right) \left(
B+Y^{-1}T\right) \sqrt{\frac{2}{N}}\sum_{k\in \mathbb{Z}}\left| \frac{2m-N}{N%
}+2k+1\right\rangle \\
&=&\frac{1}{\sqrt{2}}\phi _{2m-N}^{\left( 0,0\right) }-\frac{e^{-2\pi im/N}}{%
\sqrt{2N}}\left( \sum_{a=N/2}^{N-1}e^{-2\pi ia\left( 2m\right)
/N}-\sum_{a=N/2}^{N-1}e^{-2\pi ia\left( 2m\right) /N}\right) \\
&&\times \sum_{b\,\text{odd}}^{N-1}e^{i\pi b/N}\left( \mathcal{F}%
^{-1}\right) _{ab}\;\Phi _{b}^{(0,1/2)}
\end{eqnarray*}
Thus, 
\[
\left( \Phi _{n}^{\left( 0,0\right) },F\Phi _{m}^{\left( 0,0\right) }\right)
_{P}=\left\{ 
\begin{array}{ll}
\sum_{a=0}^{N/2-1}\left( \mathcal{F}^{N}\right) _{na}^{-1}\left( 
\begin{array}{ll}
0 & 0 \\ 
0 & \mathcal{F}^{N/2}
\end{array}
\right) _{am} & n\text{ even} \\ 
e^{i\pi \left( n-\left( 2m-N\right) \right) /N}\sum_{a=0}^{N/2-1}\left( 
\mathcal{F}^{N}\right) _{na}^{-1}\left( 
\begin{array}{ll}
0 & 0 \\ 
0 & \mathcal{F}^{N/2}
\end{array}
\right) _{am} & n\,\text{odd}
\end{array}
\right. 
\]
Combining all our previous results, we see that equation \ref{matrix} holds
for any $m$. This completes the proof of the theorem. 
\endproof%

\begin{lemma}
The matrix $B_{nm}=\left( \Phi _{n}^{(0,0)},F\Phi _{m}^{(0,0)}\right) $ is
unitary.
\end{lemma}

Proof. Consider the case $0\leq n,m<N/2$%
\begin{eqnarray*}
\sum_{j=0}^{N-1}\left( F^{\dagger }\right) _{nj}F_{jm}
&=&\sum_{j=0}^{N-1}F_{jn}^{*}F_{jm} \\
&=&\frac{1}{2}\sum_{j\text{ even}}\delta _{j,2n}\delta _{j,2m} \\
&&+\sum_{j\text{ odd}}\left( \frac{1}{N/\sqrt{2}}e^{-i\pi \left( j-2n\right)
/N}\sum_{a=0}^{N/2-1}e^{-2\pi ija/N}e^{2\pi ian/\left( N/2\right) }\right) \\
&&\times \left( \frac{1}{N/\sqrt{2}}e^{-i\pi \left( j-2n\right)
/N}\sum_{a=0}^{N/2-1}e^{-2\pi ija/N}e^{2\pi ian/\left( N/2\right) }\ \right)
\\
&=&\frac{1}{2}\delta _{n,m}+\frac{2e^{2\pi i\left( n-m\right) /N}}{N^{2}}%
\sum_{j\text{ odd}}\sum_{a,b=0}^{N/2-1}e^{2\pi ij\left( b-a\right)
/N}e^{2\pi i\left( an-bm\right) /N/2} \\
&=&\frac{1}{2}\delta _{n,m}+\frac{e^{2\pi i\left( n-m\right) /N}}{N}%
\sum_{a=0}^{N/2-1}e^{2\pi ia\left( n-m\right) /N/2} \\
&=&\delta _{n,m}.
\end{eqnarray*}
The remaining cases use an analogous calculation. We omit the details here. 
\endproof%

\section{Parity Conservation in the Quantum Dynamics}

We can see explicitly that at the fixed point $\theta =\left( 0,0\right) $
the dynamics is invariant under the symmetry $x\rightarrow 1-x$ and $%
p\rightarrow 1-p$. To see this, we define the parity operator $P$ such that 
\begin{eqnarray*}
P\left| x\right\rangle &=&\left| 1-x\right\rangle , \\
P\left| p\right\rangle &=&\left| 1-p\right\rangle \text{. }
\end{eqnarray*}
Then observe that 
\begin{eqnarray*}
PF\phi _{m}^{\left( 0,0\right) }
&=&P(E_{x}+X^{-1/2}O_{x})(B+Y^{-1}T)SP^{-1}P\phi _{m}^{\left( 0,0\right) } \\
&=&(O_{x}+X^{1/2}E_{x})(T+YB)SP\phi _{m}^{\left( 0,0\right) } \\
&=&(E_{x}+X^{-1/2}O_{x})(B+YT)SP\phi _{m}^{\left( 0,0\right) }.
\end{eqnarray*}
Since 
\[
YTSP\phi _{m}^{\left( 0,0\right) }=Y^{-1}TSPY\phi _{m}^{\left( 0,0\right)
}=Y^{-1}TSPY\phi _{m}^{\left( 0,0\right) } 
\]
we see that 
\[
PF\phi _{m}^{\left( 0,0\right) }=FP\phi _{m}^{\left( 0,0\right) } 
\]
Thus we see that on the subspace $\mathcal{H}_{\hbar }\left( 0\right) $, the
dynamics commutes with the parity operator, hence is conserved.

The authors would like to thank Andrew Lesniewski, Christopher King, Lev
Kaplan, and Eric Heller for many fruitful discussions. RR is supported by a
National Science Foundation Graduate Research Fellowship.

\end{document}